\documentclass[reprint,prd,nofootinbib]{revtex4-2}
\usepackage[caption=false]{subfig}
\usepackage{graphicx}
\usepackage{amsmath}
\usepackage{hyperref}
\usepackage{paralist}
\usepackage[]{xcolor}
\usepackage[normalem]{ulem}
\usepackage{orcidlink}

\newlength{\okinalen}
\newcommand{\okina}{\hbox to.666\okinalen{\hss`\hss}}
\pdfoutput=1
\newcommand{\Siena}{Department of Physics and Astronomy, Siena College, 515 Loudon Rd, Loudonville, NY 12211, USA}

\newcommand{\IfA}{Institute for Astronomy, University of Hawai\okina i, 2680 Woodlawn Drive, Honolulu, HI 96822, USA}

\newcommand{\UHM}{Department of Physics and Astronomy, University of Hawai\okina i, Watanabe Hall, 2505 Correa Road, Honolulu, HI 96822, USA}

\newcommand{\trgb}{TRGB}

\newcommand{\mtrgb}{$\textrm{M}_\textrm{I}^{\textrm{\trgb}}$}

\begin{document}


\newcommand{\tnm}[1]{{\tablenotemark{#1}}}
\newcommand{\tnt}[2]{{\tablenotetext{#1}{#2}}}
\newcommand{\fig}[2]{{fig.\,{#1}{#2}}}
\newcommand{\tab}[1]{{table\,{#1}}}
\newcommand{\eqn}[1]{{eq.\,{#1}}}

\newcommand{\Hawaii}{{Hawai`i}}

\newcommand{\eg}{{\it e.g.}}
\newcommand{\ie}{{\it i.e.}}
\newcommand{\etc}{{\it etc.}}
\newcommand{\etal}{{\it et~al.}}
\newcommand{\adhoc}{{\it ad~hoc}}
\newcommand{\insitu}{{\it in situ}}
\newcommand{\apriori}{{\it a~priori}}
\newcommand{\postfacto}{{\it post~facto}}

\newcommand{\half}{{\frac{1}{2}}}
\newcommand{\mean}[1]{\langle{#1}\rangle}
\newcommand{\dif}{\mathrm{d}}
\newcommand{\overbar}[1]{\mkern 1.5mu\overline{\mkern-1.5mu#1\mkern-1.5mu}\mkern 1.5mu}
\newcommand{\oforder}{\mathcal{O}}
\newcommand{\sci}[2]{{{#1}\times10^{#2}}}

\newcommand{\xbar}{\bar x}
\newcommand{\xyz}{(x,y,z)}
\newcommand{\xyzdot}{(\dot{x},\dot{y},\dot{z})}
\newcommand{\aei}{(a,e,i)}
\newcommand{\qei}{(q,e,i)}
\newcommand{\aeiH}{(a,e,i,H)}
\newcommand{\qeiD}{(q,e,i,D)}
\newcommand{\OoM}{(\Omega,\omega,M)}
\newcommand{\vo}{\vec{o}}
\newcommand{\vx}{\vec{x}}
\newcommand{\vxdot}{\vec{\dot x}}
\newcommand{\vxavg}{\mean{\vec{x}}}
\newcommand{\vy}{\vec{y}}
\newcommand{\vyavg}{\mean{\vec{y}}}
\newcommand{\vz}{\vec{z}}
\newcommand{\vzavg}{\mean{\vec{z}}}
\newcommand{\Havg}{\mean{H}}
\newcommand{\deltav}{\Delta v}
\newcommand{\node}{\Omega}
\newcommand{\aperi}{\omega}
\newcommand{\lperi}{\tilde\omega}

\newcommand{\todo}[2]{{\color{red}\bf #1 - #2}}
\newcommand{\note}[1]{{\color{blue}{\bf #1}}}
\newcommand{\XXX}{{\color{red}\bf XXX}}
\newcommand{\citepna}[1]{{{\color{red} (#1)}}}
\newcommand{\citetna}[1]{{{\color{red} #1}}}

\newcommand{\sn}{$S/N$}
\newcommand{\SN}{$S/N$}
\newcommand{\rd}{$^{rd}$}
\newcommand{\st}{$^{st}$}
\newcommand{\nd}{$^{nd}$}
\newcommand{\fManx}{{\mathrm{f}_\mathrm{Manx}}}
\newcommand{\digesttwo}{{\texttt{digest2}}}

\newcommand{\HtwoO}{{H$_2$O}}
\newcommand{\CO}{{CO}}
\newcommand{\COtwo}{{CO$_2$}}

\newcommand{\arcdeg}{{^{\circ}}}
\newcommand{\arcmin}{^{\prime}}
\newcommand{\arcsec}{^{\prime\prime}}
\newcommand{\h}{^\mathrm{h}}
\newcommand{\m}{^\mathrm{m}}
\newcommand{\s}{^\mathrm{s}}
\newcommand{\Mpc}{\,\mathrm{Mpc}}
\newcommand{\kpc}{\,\mathrm{kpc}}
\newcommand{\pc}{\,\mathrm{pc}}
\newcommand{\au}{\,\mathrm{au}}
\newcommand{\km}{\,\mathrm{km}}
\newcommand{\kph}{\,\mathrm{km}/\mathrm{h}}
\newcommand{\kps}{\,\mathrm{km}\,\mathrm{s}^{-1}}
\newcommand{\ft}{\,\mathrm{ft}}
\newcommand{\meter}{\,\mathrm{m}}
\newcommand{\cm}{\,\mathrm{cm}}
\newcommand{\mm}{\,\mathrm{mm}}
\newcommand{\um}{\,\mu \mathrm{m}}
\newcommand{\nm}{\,\mathrm{nm}}
\newcommand{\rad}{\,\mathrm{rad}}
\newcommand{\rms}{\,\mathrm{(rms)}}
\newcommand{\anno}{\,\mathrm{a}}
\newcommand{\yr}{\,\mathrm{yr}}
\newcommand{\Myr}{\,\mathrm{Myr}}
\newcommand{\Gyr}{\,\mathrm{Gyr}}
\newcommand{\Day}{\,\mathrm{day}}
\newcommand{\days}{\,\mathrm{d}}
\newcommand{\dayperyear}{\,\mathrm{d}/\mathrm{yr}}
\newcommand{\vrk}{\,\mathrm{vrk}}
\newcommand{\degrees}{\,\mathrm{deg}}
\newcommand{\hours}{\,hours}
\newcommand{\hour}{\,\mathrm{h}}
\newcommand{\hourperday}{\,\mathrm{h}/\mathrm{d}}
\newcommand{\minute}{\,\mathrm{min}}
\newcommand{\second}{\,\mathrm{s}}
\newcommand{\mps}{\,\meter\,\second^{-1}}
\newcommand{\Hz}{\,\mathrm{Hz}}
\newcommand{\mags}{\,\mathrm{mag}}
\newcommand{\K}{\,\mathrm{K}}
\newcommand{\J}{\,\mathrm{J}}
\newcommand{\N}{\,\mathrm{N}}
\newcommand{\kg}{\,\mathrm{kg}}
\newcommand{\g}{\,\mathrm{g}}
\newcommand{\AMU}{\,\mathrm{AMU}}
\newcommand{\W}{\,\mathrm{W}}
\newcommand{\MW}{\,\mathrm{MW}}
\newcommand{\degC}{\arcdeg\mathrm{C}}
\newcommand{\degK}{\mathrm{K}}
\newcommand{\Jy}{\,\mathrm{Jy}}
\newcommand{\mJy}{\,\mathrm{mJy}}
\newcommand{\Mearth}{\,\mathrm{M}_\oplus}

\newcommand{\asteroid}[2]{{({#1})\,{#2}}}
\newcommand{\designation}[2]{{{#1}\,{#2}}}
\newcommand{\TC}{{2008\,TC$_3$}}
\newcommand{\RH}{{2006\,RH$_{120}$}}
\newcommand{\Bennu}{{(101955)\,Bennu}}

\newcommand{\Sthree}{{C/2014$\,$S3$\,$PANSTARRS}}
\newcommand{\Uone}{{1I/2017 U1 (‘Oumuamua)}}


\newcommand{\gps}{\ensuremath{g_{\rm P1}}}
\newcommand{\rps}{\ensuremath{r_{\rm P1}}}
\newcommand{\ips}{\ensuremath{i_{\rm P1}}}
\newcommand{\zps}{\ensuremath{z_{\rm P1}}}
\newcommand{\yps}{\ensuremath{y_{\rm P1}}}
\newcommand{\wps}{\ensuremath{w_{\rm P1}}}
\newcommand{\grizy}{\gps\rps\ips\zps\yps}
\newcommand{\JHK}{\ensuremath{JHK}}
\newcommand{\V}{\ensuremath{V}}

\newcommand{\PS}{\protect \hbox {Pan-STARRS}}
\newcommand{\PSone}{\protect \hbox {Pan-STARRS1}}
\newcommand{\PStwo}{\protect \hbox {Pan-STARRS2}}
\newcommand{\PSfour}{\protect \hbox {Pan-STARRS4}}
\newcommand{\knownserver}{{\tt known\_server}}


\newcommand\aj{AJ}
\newcommand\psj{{PSJ}}
\newcommand\araa{{ARA\&A}}
\renewcommand\apj{{ApJ}}
\newcommand\apjl{{ApJL}}     
\newcommand\apjs{{ApJS}}
\renewcommand\ao{{ApOpt}}
\newcommand\apss{{Ap\&SS}}
\newcommand\aap{{A\&A}}
\newcommand\aapr{{A\&A~Rv}}
\newcommand\aaps{{A\&AS}}
\newcommand\azh{{AZh}}
\newcommand\baas{{BAAS}}
\newcommand\icarus{{Icarus}}
\newcommand\jaavso{{JAAVSO}}  
\newcommand\jrasc{{JRASC}}
\newcommand\memras{{MmRAS}}
\newcommand\mnras{{MNRAS}}
\renewcommand\pra{{PhRvA}}
\renewcommand\prb{{PhRvB}}
\renewcommand\prc{{PhRvC}}
\renewcommand\prd{{PhRvD}}
\renewcommand\pre{{PhRvE}}
\renewcommand\prl{{PhRvL}}
\newcommand\pasp{{PASP}}
\newcommand\pasj{{PASJ}}
\newcommand\qjras{{QJRAS}}
\newcommand\skytel{{S\&T}}
\newcommand\solphys{{SoPh}}
\newcommand\sovast{{Soviet~Ast.}}
\newcommand\ssr{{SSRv}}
\newcommand\zap{{ZA}}
\renewcommand\nat{{Nature}}
\newcommand\iaucirc{{IAUC}}
\newcommand\aplett{{Astrophys.~Lett.}}
\newcommand\apspr{{Astrophys.~Space~Phys.~Res.}}
\newcommand\bain{{BAN}}
\newcommand\fcp{{FCPh}}
\newcommand\gca{{GeoCoA}}
\newcommand\grl{{Geophys.~Res.~Lett.}}
\renewcommand\jcp{{JChPh}}
\newcommand\jgr{{J.~Geophys.~Res.}}
\newcommand\jqsrt{{JQSRT}}
\newcommand\memsai{{MmSAI}}
\newcommand\nphysa{{NuPhA}}
\newcommand\physrep{{PhR}}
\newcommand\physscr{{PhyS}}
\newcommand\planss{{Planet.~Space~Sci.}}
\newcommand\procspie{{Proc.~SPIE}}

\newcommand\actaa{{AcA}}
\newcommand\caa{{ChA\&A}}
\newcommand\cjaa{{ChJA\&A}}
\newcommand\jcap{{JCAP}}
\newcommand\na{{NewA}}
\newcommand\nar{{NewAR}}
\newcommand\pasa{{PASA}}
\newcommand\rmxaa{{RMxAA}}

\newcommand\maps{{M\&PS}}
\newcommand\aas{{AAS Meeting Abstracts}}
\newcommand\dps{{AAS/DPS Meeting Abstracts}}

\preprint{}

\title{Tip of the Red Giant Branch Bounds on the Neutrino Magnetic Dipole Moment Revisited}

\author{Noah Franz \orcidlink{0000-0003-4537-3575}}
\affiliation{\IfA}
\affiliation{\Siena}

\author{Mitchell Dennis \orcidlink{0000-0001-9066-0552}}
\affiliation{\IfA}

\author{Jeremy Sakstein \orcidlink{0000-0002-9780-0922}}
\affiliation{\UHM}

\begin{abstract}
    We use a novel method to constrain the neutrino magnetic dipole moment ($\mu_{\nu}$) using the empirically-calibrated tip of the red giant branch I-band magnitude that fully accounts for uncertainties in stellar physics.~Our method uses machine learning to emulate the results of stellar evolution codes.~This reduces the I-Band magnitude computation time to milliseconds, which enables a Bayesian statistical analysis where $\mu_{\nu}$ is varied simultaneously with the stellar physics, allowing for a complete exploration of parameter space.~We find the region $\mu_{\nu} \leq 6\times10^{-12}\mu_{\textrm{B}}$ (with $\mu_{\textrm{B}}$ the Bohr magneton), previously believed to be excluded, is unconstrained after accounting for degeneracies with  stellar physics.~It is likely that larger values are similarly unconstrained.~We discuss the implications of our results for future neutrino magnetic dipole moment searches and for other astrophysical probes.
\end{abstract}

\date{\today}

\maketitle

\section{Introduction}\label{sec:intro}

The high temperatures and densities in the cores of stars makes them  naturally occurring laboratories that allow astronomers to probe the properties of light ($<10$keV), weakly interacting particles (e.g., \cite{Raffelt1996Book, Raffelt2000}).~Examples of theories that can be probed using these stellar laboratories are those that include axions \cite{Marsh2017, Luzio2020, Chadha-Day2022}, hidden photons \cite{Fabbrichesi2021}, and a larger neutrino magnetic dipole moment (MDM) than predicted by the Standard Model of Particle Physics (SM) \cite{Broggini2012, alexander2016}.~Unlike terrestrial laboratories, stellar interiors are subject to large uncertainties due to environmental variations (e.g., mass and metallicity) and uncertain input physics e.g., mixing length, opacity, and nuclear reaction rates.~Any attempt to bound new physics using these objects should simultaneously vary the stellar physics parameters to incorporate their effects (including degeneracies) into the final uncertainty.~Unfortunately, the long run-times of stellar structure codes ($\sim$hours or longer) prevents the use of Bayesian statistical methods such as Markov Chain Monte Carlo (MCMC) sampling, which require run-times of order seconds or shorter.~For this reason, previous works have either held the stellar physics fixed or varied parameters individually with all others held fixed.~

 Recently, two of us presented a novel method that enables MCMC analyses of stellar observations to constrain stellar input physics \cite{Dennis:2023ldw} and the axion-electron coupling \cite{Dennis:2023kfe}.~This method trains a machine learning emulator to predict stellar observables as a function of the stellar and beyond SM physics parameters.~The emulator, which evaluates in milliseconds, is called by the MCMC sampler at each point in parameter space rather than a stellar structure code, allowing for rapid convergence.~Applying this method to tip of the red giant branch (TRGB) bounds on the axion-electron coupling revealed that the range $0\le\alpha_{26}\le 2$, previously believed to be excluded, is unconstrained once the degeneracies between stellar parameters are accounted for \cite{Dennis:2023kfe}.~In this work, we  apply this method to reevaluate TRGB bounds on the neutrino MDM, which affects stars in a similar manner to light axions.~The SM predicts a neutrino MDM $\mu_\nu = 3.20 \times 10^{-19} \mu_B$, where $\mu_B=e/2m_e$ is the Bohr Magneton \citep{Mohapatra1991, Winter1991, Raffelt2000};~too small to effect stellar evolution.~In contrast, some extensions of the SM predict neutrino MDMs that are larger by several orders of magnitude \cite{2014PhRvD..89e5009A}.~By constraining the neutrino MDM we can therefore eliminate extensions of the SM.

TRGB stars provide the most stringent constraints on the neutrino MDM \cite{francesco&raffelt2020}.~Low-mass stars ($M\lesssim2.25{\rm M}_{\odot}$ depending on metallicity and other parameters) enter this stage of evolution after reaching a core temperature of $\approx10^8$K in a degenerate helium core surrounded by a hydrogen burning shell.~At this temperature, a runaway helium reaction creates an explosion resulting in the \textit{helium flash}.~The star rapidly moves to the horizontal branch, leaving a visible discontinuity in the color-magnitude diagram --- the TRGB.~The \trgb\ consistency across parameter space (e.g.~initial helium abundance, initial metallicity) in the  I-band magnitude,~\mtrgb, makes it useful as a standard candle \citep{Shapley_1930, Baade_1944, DaCosta_Armandroff_1990, Freedman&Madore2010, Freedman&Madore2020}.~\mtrgb\ and its uncertainty are calculated either by calibrating observational data \citep{francesco&raffelt2020, Lee1993} or by calculating it theoretically  using stellar structure codes \citep{serenelli2017, Ippocratis2022}.~The error on the observational calibration of \mtrgb\ is smaller than that on the theoretical calculation because of uncertainties in the stellar input physics \cite{serenelli2017, Ippocratis2022} and large empirical errors on the bolometric corrections needed to convert the outputs of stellar structure codes to magnitudes.~For this reason, observational calibration of \mtrgb\ is the standard in the literature.

 A larger neutrino MDM than predicted by the SM causes additional energy losses in the plasma and pair-production channels \citep{heger2008} that  decrease the core temperature.~In turn, the hydrogen shell burning phase is lengthened, and more helium  is deposited on the core, resulting in a  more energetic explosion  and consequentially a brighter Helium flash \citep{Raffelt1990, Raffelt&Weiss1992, francesco&raffelt2020}.~Many works have  used this effect to constrain the  neutrino MDM  by comparing empirical calibrations of ~\mtrgb\ with the results of incorporating the additional energy losses due to a large MDM into stellar structure codes \citep{Raffelt&Weiss1992, Castellani&degl'Innocenti1993, Raffelt2000, heger2008, francesco&raffelt2020}.~Most recently, \cite{francesco&raffelt2020} obtained the most restrictive limit:~$\mu_\nu < 1.2 \times10^{-12} \mu_B$ at the 95\% confidence level using calibrated observations of $\omega$-Centauri.~In this work, we reanalyze the calibrations studied by \cite{francesco&raffelt2020} using our  MCMC-machine learning method to account for degeneracies with the mass, metallicity, and initial helium abundance, and find that the region $\mu_\nu\leq6\times10^{-12}\mu_B$ is unconstrained once these degeneracies are accounted for.

This paper is organized as follows:~\S\ref{sec:method} outlines the  methodology we implement to constrain the neutrino MDM including subsections on the modification of the stellar evolution code, our machine learning emulator, and our MCMC analysis.~\S\ref{sec:results} presents and discusses the results of the stellar evolution code modifications, the machine learning emulator, and the MCMC analysis.~\S\ref{sec:conclusion}  discusses the  implications of our results, examines the limitations of our methodology, and suggests future work.~We conclude in \S\ref{sec:real_conclusions}.

All of the codes used in this work including our MESA inlists, modifications, and output files;~machine learning emulators;~and MCMC scripts are available in a reproduction package at the following URL: \href{https://doi.org/10.5281/zenodo.8173321}{https://doi.org/10.5281/zenodo.8173321} \cite{franz_noah_2023_8173321}.

\section{Method}
\label{sec:method}

\subsection{Stellar Evolution Code and Training Set}
\label{sec:mesa}

  We modified the stellar evolution code MESA version 12778 \citep{Paxton2011, Paxton2013, Paxton2015, Paxton2018, Paxton2019,Jermyn:2022mst} to add an extra component to the SM neutrino energy loss, giving a total loss rate per unit mass:
\begin{align}
\varepsilon_{\textrm{tot}} = \varepsilon_\textrm{plasma}^\mu + \varepsilon_\textrm{pair}^\mu + \varepsilon_{\textrm{SM}},~\label{eq:eTot}
\end{align}
where $\varepsilon_{\textrm{tot}}$ is the energy loss per unit mass, $\varepsilon_{\textrm{SM}}$ is the SM energy loss rate calculated by MESA, $\varepsilon_\textrm{plasma}^\mu$ is the plasma energy loss rate due to neutrino MDM given by \citep{heger2008, Haft1994},
\begin{align}
    \varepsilon_\textrm{plasma}^\mu &= 0.318~ \varepsilon_\textrm{plasma}~\left(\frac{10~\textrm{keV}}{\omega_\textrm{plasma}}\right)^2~\mu_{12}^2, \label{eq:ePlasma}
\end{align}
and $\varepsilon_\textrm{pair}$ is the pair production energy loss rate due to neutrino MDM given by \citep{heger2008},
\begin{align}
    \varepsilon_\textrm{pair}^\mu &= 1.6\times 10^{11} \textrm{erg}~\textrm{g}^{-1}~\textrm{s}^{-1} \times \frac{\mu_{10}^2}{\rho_4}~\textrm{e}^{-118.5/T_8}.~\label{eq:ePair}
\end{align}
In equation \eqref{eq:ePlasma}, $\varepsilon_\textrm{plasma}$ is the SM plasma energy loss rate, $\mu_{12} = \mu_\nu/10^{12}~\mu_B$, where $\mu_\nu$ is the nuetrino MDM, and $\omega_\textrm{plasma}$ is given by,
\begin{align}
    \omega_\textrm{plasma}^2 = \frac{4\pi\alpha n_e}{m_e}, \label{eq:omega}
\end{align}
where $\alpha$ is the fine-structure constant, $n_e$ is the electron number density, and $m_e$ is the mass of an electron.~In equation \eqref{eq:ePair}, $\mu_{10} = \mu_\nu/10^{10} ~\mu_B$, $\rho_4 = \rho/10^{4} ~\textrm{g}~\textrm{cm}^{-3}$ where $\rho$ is the density, and $T_8 = T/10^{8}\textrm{K}$ where $T$ is the temperature.

In order to directly compare our simulated results with observations, we must convert from MESA output quantities to \mtrgb.~Furthermore, the empirical calibration of 
 \mtrgb\ has a strong $(V-I)$ color dependence.~Any particular MESA model could have a final $(V-I)$ which differs from the reference $(V-I)$ color (zero-point) of the given calibration.~It is therefore necessary to use a color-correction to shift the model output (both $(V-I)$ and \mtrgb) to the  reference color.~The color dependence of each system is unique, leading to different color-corrections for each system.~We used the empirical bolometric corrections of Worthey \& Lee (WL) \cite{worthey&lee2011} to make this conversion.~WL takes the surface gravity, [Fe/H], luminosity, and effective temperature from the MESA models and calculates bolometric corrections used to calculate the I-band magnitude, $(V-I)$ color, and both the empirical I-band and $(V-I)$ color errors.~These were then passed into the color-correction of the form \mtrgb$ = f(M_{I}, (V-I))$ given in Table \ref{tab:calib}.~The procedure above is tantamount to assuming that the TRGB I-band magnitude is due solely to the brightest star.~Whereas this is a reasonable first-approximation \cite{Viaux2013}, we discuss how one could use our ML emulator to make a more realistic theoretical prediction in \S\ref{sec:limits}.~The purpose of our present study is to compare with previous bounds on $\mu_\nu$, which make the same single-star approximation \cite{Viaux2013,Capozzi2020}, so adopting this approximation ensures that our conclusions are driven by uncertainties and degeneracies in the stellar modeling parameters and not by difference in the theoretical prescriptions.

Using the modified version of MESA and the WL conversion, we ran a grid of 146,250  MESA models from the pre-main-sequence to the TRGB (defined as the point where the power in helium burning exceeds $10^6$ ergs/s) with varying stellar mass ($M$), metallicity ($Z$), initial helium fraction ($Y$), and  neutrino MDM ($\mu_{12}$).~The grid includes input ranges of $0.7 \leq M/{\rm M}_\odot\leq2.25$, $0.2\leq Y\leq0.3$, $10^{-5}\leq Z\leq0.04$, and $0 \leq \mu_{12} \leq 6$.~The range of $M$, $Y$, and $Z$ covers the range where a core helium flash is expected \cite{2013sse..book.....K}, and the range of $\mu_{12}$ covers values where \mtrgb\ does not differ from the SM prediction to values where main-sequence stars begin to be affected.~The ranges of $M$, $Y$, and $Z$ were  chosen to be larger than the range where a core helium flash is expected in order to account for the possibility that including a neutrino MDM could cause a core flash in models that would not contribute to the TRGB in the SM, and to ensure that the boundary between a core flash and stable helium burning is resolved.~As we will discuss shortly, this is essential for ensuring  accurate emulation.~We used 15, 10, 25, and 38 points for M, Y, Z, and $\mu_{12}$ respectively.~The spacing in $M$ and $Y$ was linear and the spacing in $Z$ and $\mu_{12}$ was logarithmic.~The training data was generated uniformly across parameter space (either linearly or logarithmically);~however we note that as the dimensionality of  the parameter space increases, more efficient sampling methods such as Latin hypercube sampling will be necessary.~Due to limited computation time, we did not vary other parameters predicted to have an effect on the brightness of the \trgb\ such as mixing length, mass loss, opacity, and neutrino loss rate \cite{serenelli2017,Ippocratis2022}.~Simultaneously varying these parameters can only add more uncertainties and can therefore only strengthen our conclusions.~All MESA parameters that we hold constant assume identical values to to those in \cite{DennisSakstein2023}, with the exception of the mixing length, which was set to $\alpha_{MLT}=2.0$.~The precise choices we made for these parameters can be found in the MESA inlists in our reproduction package \cite{franz_noah_2023_8173321}.

\subsection{Machine Learning Emulator}\label{sec:ML}

 We trained a machine learning (ML) algorithm to predict the I-Band magnitude, $M_I$, the $(V-I)$ color, and the empirical uncertainties on both  as a function of the MESA inputs:~$M$, $Y$, $Z$, and $\mu_{12}$.~To emulate MESA's evolution and the WL conversion calculations, we used both a classifier and a regression Deep Neural Network (DNN).~For both ML models we  used the \texttt{Tensorflow Keras} package \cite{tf, keras}.

For some choices of parameters,  the resultant stellar model did not contribute to the TRGB.~Some models did not reach the TRGB within the current age of the universe, and others burned helium stably in their core and exhibited a shell flash.~Removing these models from the training set by hand can bias the emulator  towards predicting viable models where there are none;~so we instead  trained a  DNN classifier to predict whether a given set of parameters will contribute to the TRGB or not and assigned zero likelihood to models in the latter  class when performing the MCMC.~For classification purposes, we placed MESA evolution models into three categories:~models that reached the \trgb, models that helium shell flashed\footnote{ Models were flagged as shell flashing if the post-main-sequence helium-4 mass fraction fell below 0.9 as this implies the core has begun to burn helium stably.}, and models that did not  core flash within the age of the Universe  (13.77 billion years).~A small number of models failed to converge in MESA before they reached the \trgb.~As their number was not sufficiently large to affect the training, we removed these models before beginning ML training and allowed the ML algorithm to fill in the holes in the parameter space.

ML classification algorithms perform best when the training data is relatively proportional between the respective classes.~For our data, this  was not the case.~To account for this, we  resampled the training data using the Synthetic Minority Oversampling Technique (SMOTE) \citep{smote}.~SMOTE increases the sample size of undersampled areas in the parameter space without biasing the training data.~The classifier was trained with the Adaptive Moment Estimation (ADAM) optimizer \citep{ADAM} and used categorical crossentropy loss to measure the success of the model in training.~We hand-tuned the hyperparameters to optimize the classification algorithm.

We  trained a DNN regressor to predict $M_I$, the $(V-I$) color, and the empirical uncertainties on both  as a function of $M$, $Y$, $Z$, and $\mu_{12}$.~The regressor was only trained on models that would not be ruled out by the classifier as  older than the age of the universe or helium shell flashing.~We trained the regressor using the ADAM optimizer, and used mean-squared error loss to measure the success of the regression model.~Just as we did with the classifier, we hand tuned the hyperparameters to optimize the regression algorithm.~We calculated the uncertainty associated with the regressor and, even though this error is sub-dominant to the WL errors, it must be accounted for.~Using the testing data, distributions of error were created by taking the residual of the true and predicted values.~The residuals for the ML predictions for $M_I$, $(V-I)$ color, and their associated empirical error are shown in Figure \ref{fig:errorHists}.

\newcommand{\fw}{0.93}
\begin{figure*}
\begin{center}
\qquad
\subfloat[]{
  \includegraphics[width=\fw\columnwidth]{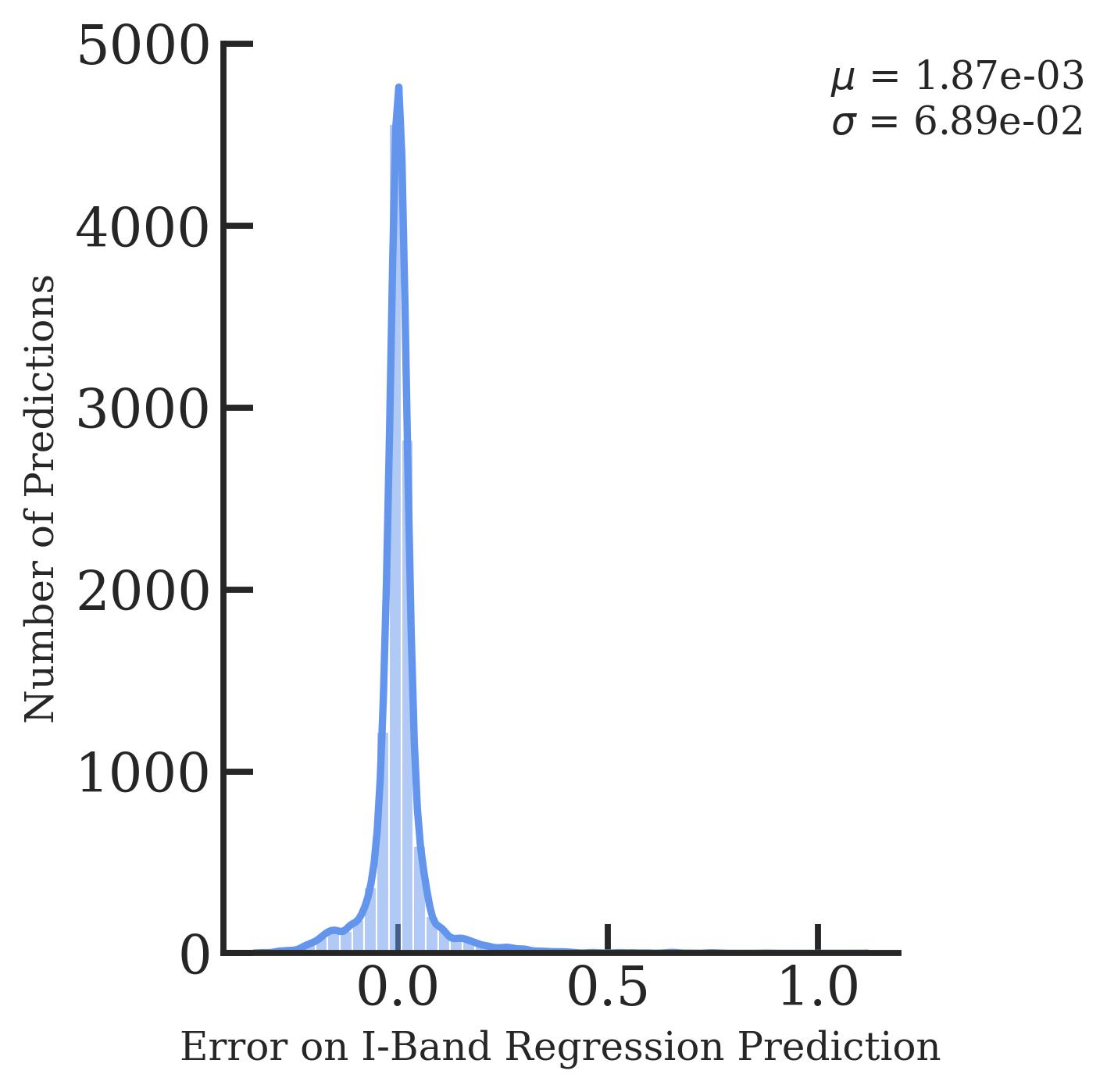}
  \label{fig:errorI}
  }
\qquad
\subfloat[]{
  \includegraphics[width=\fw\columnwidth]{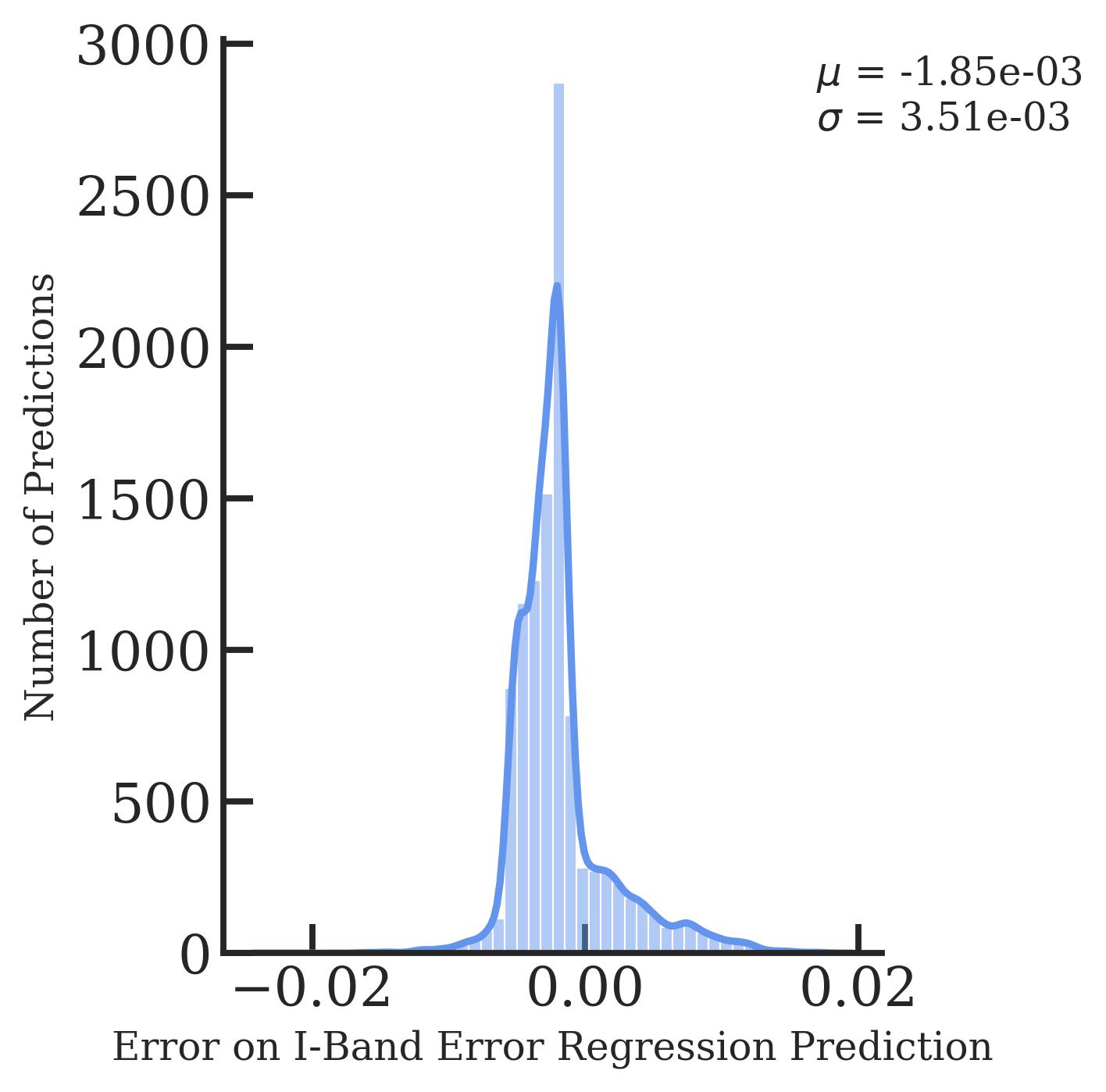}
  \label{fig:errorErrI}
  }
\qquad
\subfloat[]{
  \includegraphics[width=\fw\columnwidth]{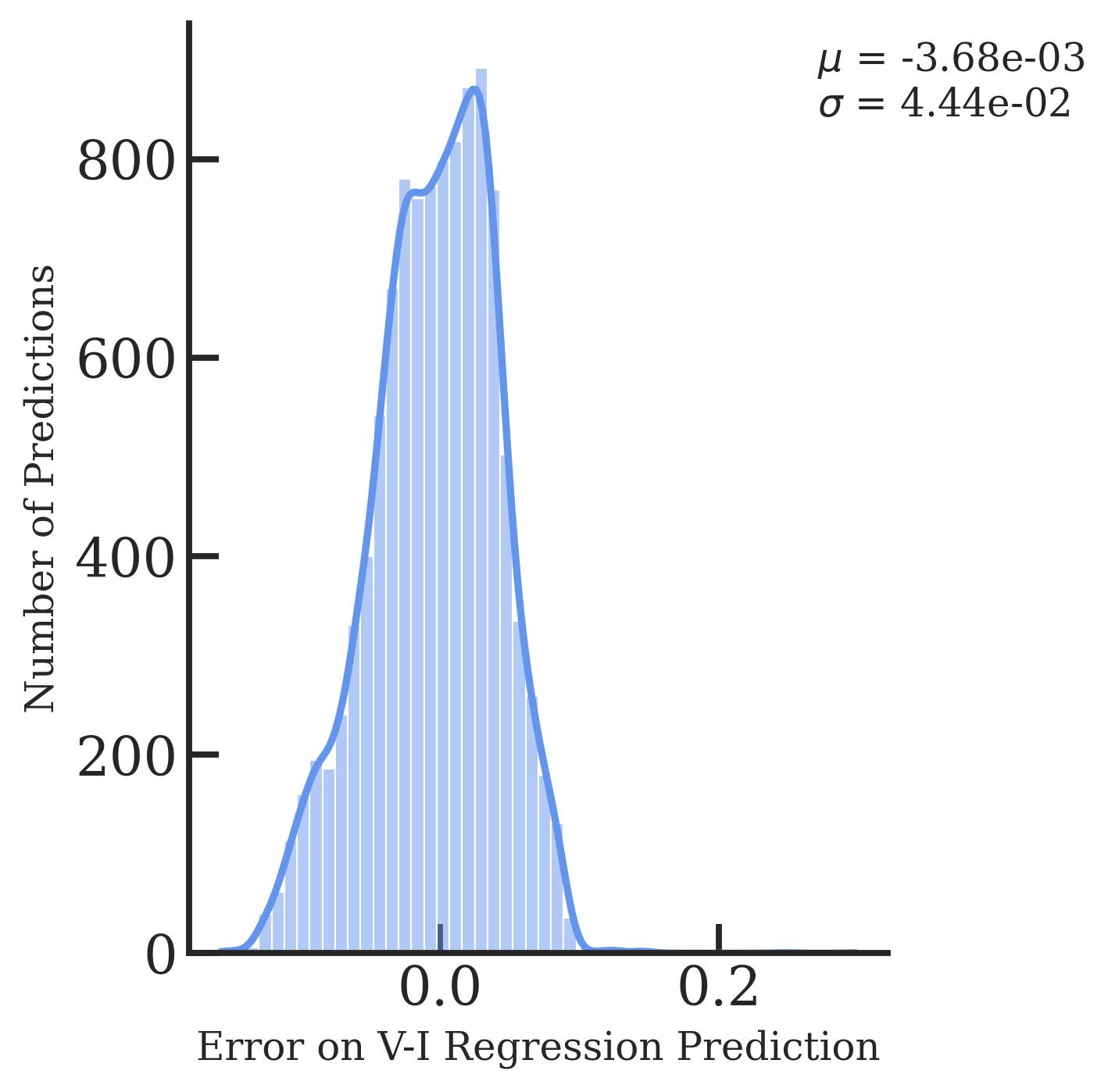}
  \label{fig:errorVI}
  }
\qquad
\subfloat[]{
  \includegraphics[width=\fw\columnwidth]{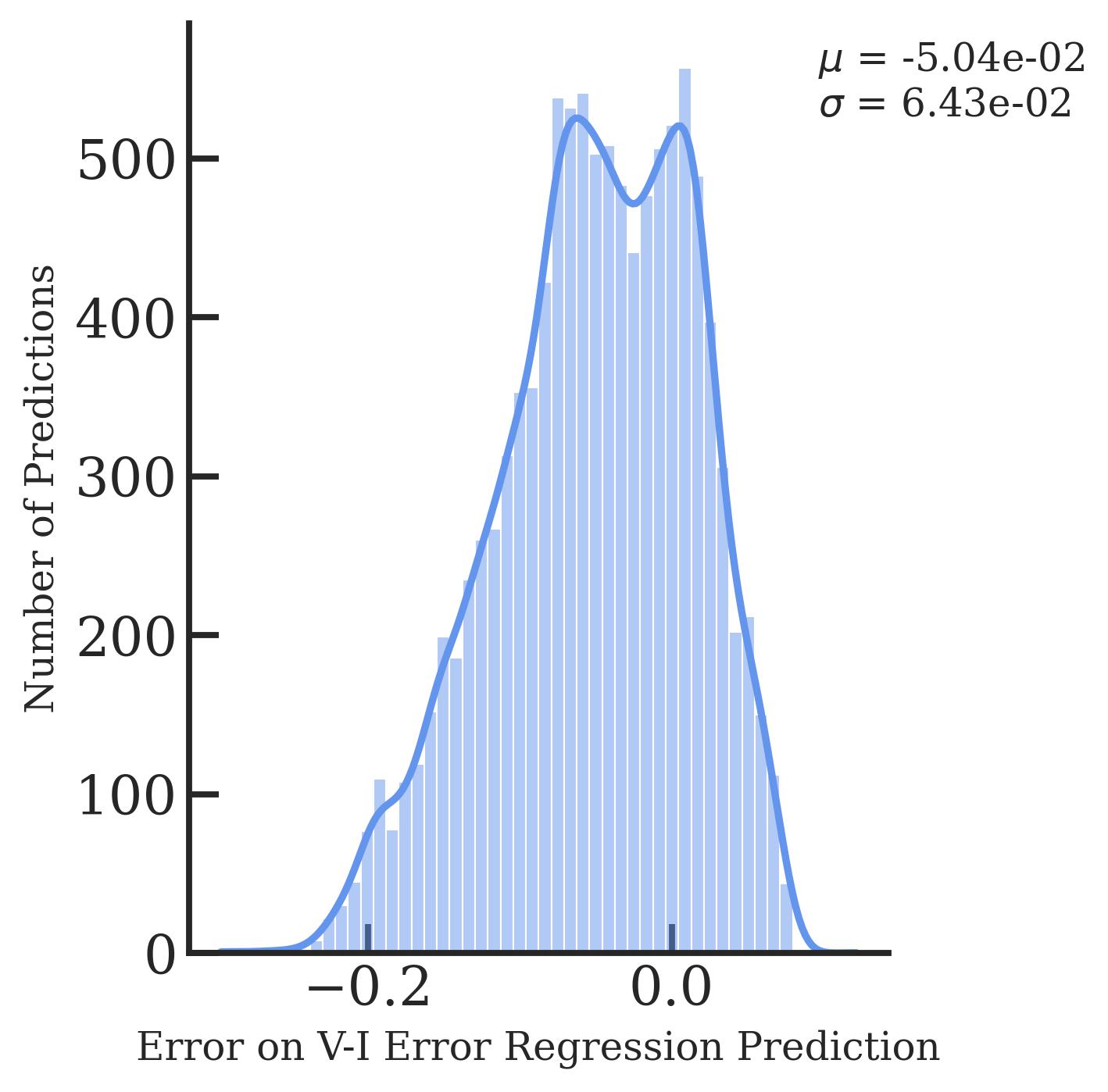}
  \label{fig:errorErrVI}
  }
\caption{Error distributions for the machine learning emulator outputs.~Figure \ref{fig:errorI} shows the error on the \mtrgb\ prediction, Figure \ref{fig:errorErrI} shows the error on the \mtrgb\ WL error prediction, Figure \ref{fig:errorVI} shows the error on the $V-I$ color prediction, and Figure \ref{fig:errorErrVI} shows the error on the $V-I$ color error prediction.~Each of these were calculated by subtracting the true value from the predicted value.~$\mu$ refers to the mean and $\sigma$ the standard deviation.
}
\label{fig:errorHists}
\end{center}
\end{figure*}

\subsection{MCMC}
\label{sec:mcmc}
We  used an MCMC sampler to attempt to constrain $\mu_{12}$.~This  compared the theoretical values of \mtrgb\ calculated using our ML emulator with the observations from the LMC, NGC 4258, and $\omega$-Centauri.

We  performed the MCMC analysis using the Python package \texttt{emcee} \citep{emcee} with the ML stellar evolution emulator to constrain the Neutrino MDM with uncertainty.~We conservatively assumed no knowledge of the input parameters and therefore used uniform priors  for all of the input parameters.~Our classification algorithm  was used to assign a probability of zero to models which do not helium flash, or that are older than the age of the universe.~Otherwise, the likelihood of a given model  was calculated using the Gaussian log-likelihood function
\begin{align}
    P(\theta~|~X) = -\frac{1}{2}~\Bigg[\frac{\left(M_{\textrm{I, obs}} - M_{\textrm{I, corr}}\right)^2}{\sigma^2} \Bigg.~+ \Bigg.\ln{\left(2\pi\sigma^2\right)}\Bigg],~\label{eq:likelihood}
\end{align}
where $M_{\textrm{I, obs}}$ is the observed value for the I-band magnitude, $M_{\textrm{I, corr}}$ is the color corrected ML prediction for the I-band magnitude given the input parameters (the color corrections for each system are given in Table \ref{tab:calib}).~In Equation \eqref{eq:likelihood}, $\sigma^2$ is given by,
\begin{align}
    \sigma^2 = \sigma_{obs}^2 +\sigma_{C}^2, \label{eq:error}
\end{align}
where $\sigma_{obs}$ is the I-band observational uncertainty and $\sigma_{C}$ is the error on the color-correction due to errors in the predicted $M_I$ and $(V-I)$, which is given by,
\begin{multline}
    \sigma_{C} \approx \left\vert\frac{\partial C}{\partial M_{I}}\right\vert^2 \sigma_I^2 + \left\vert\frac{\partial C}{\partial (V-I)}\right\vert^2 \sigma_{(V-I)}^2 + 
    \\ 2 \left\vert\frac{\partial C}{\partial M_I}\right\vert \left\vert\frac{\partial C}{\partial (V-I)}\right\vert  \sigma_I^2 ~\sigma_{(V-I)}^2~ \rho_{M_I (V-I)}, \label{eq:errProp}
\end{multline}
where $C$ is the corresponding color-correction, $M_I$ is the I-band magnitude, $(V-I)$ is the $V-I$ color, $\sigma_I$ is the predicted WL error on $M_I$, and $\sigma_{(V-I)}$ is the predicted WL error on the $V-I$ color.

 To account for the uncertainties in the ML regression algorithms predictions' for the $M_I$, $(V-I)$, and corresponding uncertainties we use a similar approach to \cite{McClintock_2019}.~We  added a randomly sampled value from each of the ML error distributions in Figure \ref{fig:errorHists} to the predicted WL error  from Equation \ref{eq:errProp}.~We  did not account for the error on the ML classifier because it  was insignificant compared to the  error on the regression algorithm.~The uncertainty  was dominated by the WL calibration and the degeneracies, not the ML algorithm.~Even  so, propagating the error in the ML  allowed us to completely characterize the uncertainties on the neutrino MDM constraint.

~We  determined that the MCMC converged if the autocorrelation time  was less than 1\% the length of the chain and has changed by less than 1\% over the last 5,000 steps.~Using this method, we consistently  found that the MCMC converged in less than 200,000 steps but we  allowed it to continue for 500,000 steps as a precaution, and to confirm that the walker is no longer influenced by its starting location.~We discarded half the samples as burn-in.~We performed our analysis on the three\footnote{Reference \cite{francesco&raffelt2020} studied a second LMC calibration reported by \cite{Freedman2020}.~Unfortunately, the color-correction is only reported over a narrow range of $(V-I)$.~Many of our models have $(V-I)$ outside this range, so we are unable to use this calibration.~There is no reason to expect that our conclusions would change were a complete color-correction available.}  calibrations of \mtrgb\ studied by \cite{francesco&raffelt2020}  given in Table \ref{tab:calib}.

\begin{table*}[]
    \centering
    \begin{tabular}{l l c c}
         \hline
         Target & \mtrgb~ Calibration~[mag]  & Color Correction & Reference\\
         \hline \hline
         LMC & $-3.958 \pm 0.046$ & $0.091[(V - I) - 1.5]^2 - 0.007[(V - I) - 1.5]$ & \cite{Yuan2019} \\ 
         NGC\,4258 & $-4.027 \pm 0.055$ & $0.091[(V - I) - 1.5]^2 - 0.007[(V - I) - 1.5]$& \cite{JangLee} \\
         $\omega$-Centauri & $-3.96 \pm 0.05$ & $-0.046(V - I) + 0.08(V - I)
^2$ & \cite{Bellazzini2001} \\
         \hline
    \end{tabular}
    \caption{Empirical \mtrgb calibrations used in this work.}
    \label{tab:calib}
\end{table*}

\section{Results}
\label{sec:results}

\subsection{MESA Output}
\begin{figure*}
\begin{center}
\qquad
\subfloat[]{
  \includegraphics[width=\fw\columnwidth]{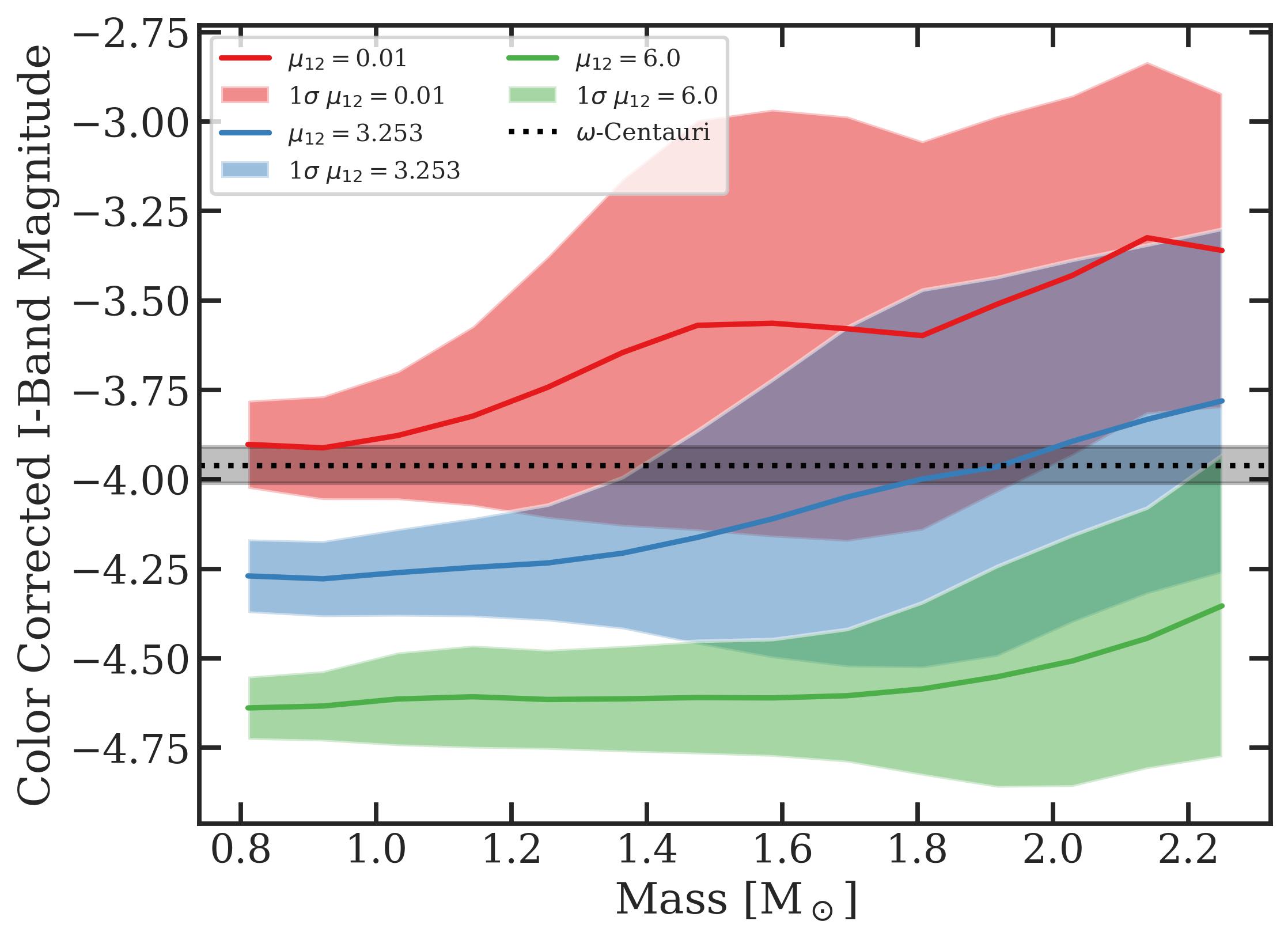}
  \label{fig:massBin}
  }
\qquad
\subfloat[]{
  \includegraphics[width=\fw\columnwidth]{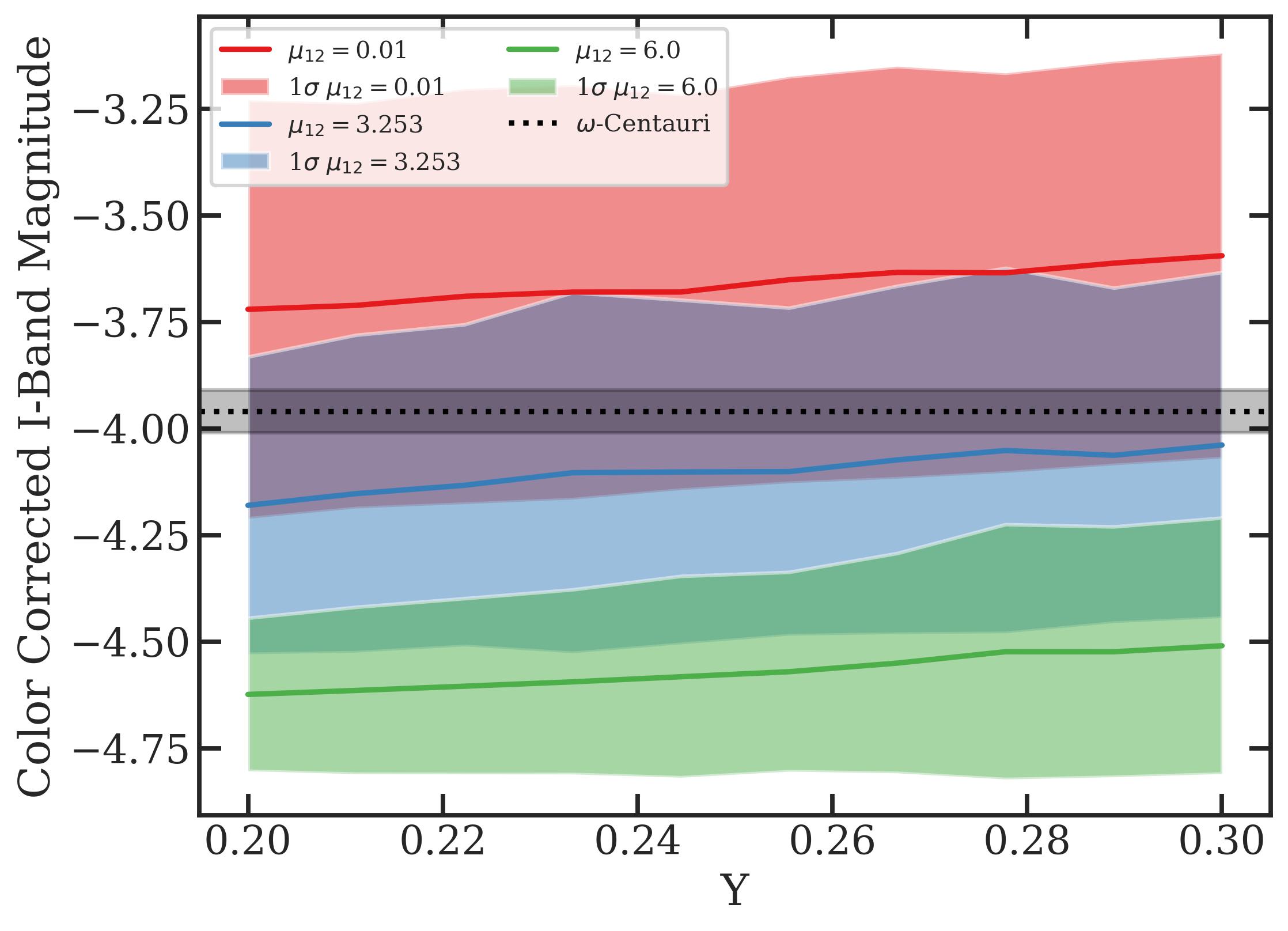}
  \label{fig:yBin}
  }
\qquad
\subfloat[]{
  \includegraphics[width=\fw\columnwidth]{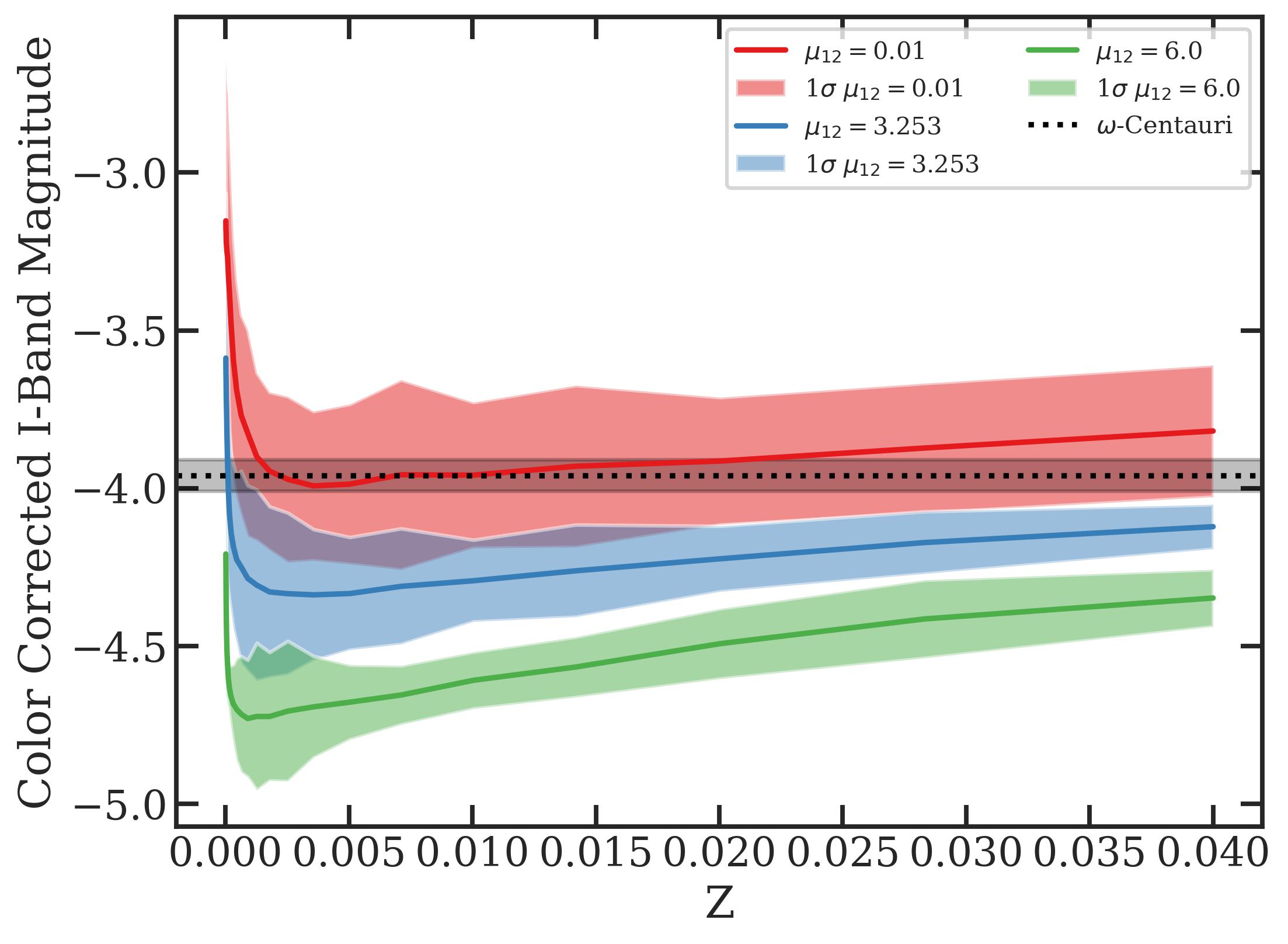}
  \label{fig:zBin}
  }
\qquad
\subfloat[]{
  \includegraphics[width=\fw\columnwidth]{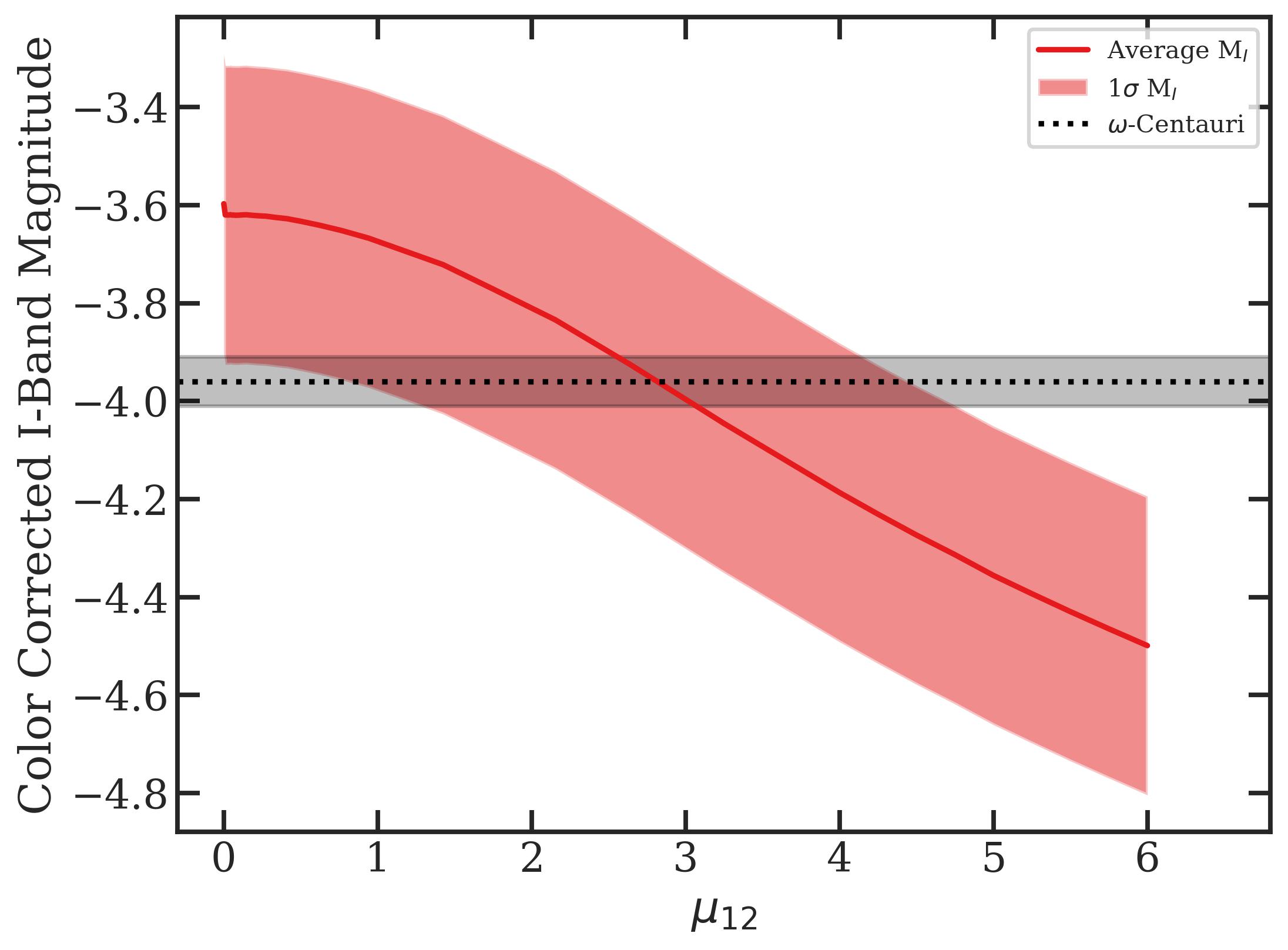}
  \label{fig:muBin}
  }
\caption{I-Band magnitude as a function of each input parameter.~Figures \ref{fig:massBin}, \ref{fig:yBin}, and \ref{fig:zBin} have three different values of $\mu_{12}$ as shown in the legend.~The grey shaded region represents an observed value of \mtrgb\ calibrated $\omega$-Centauri \cite{Bellazzini2001}.~Each point is an average across a bin of the input parameter and each error bar represents the standard deviation in that bin.}
\label{fig:binnedPlots}
\end{center}
\end{figure*}

Results from the grid exemplifying the relationship between \mtrgb\ and $M$, $Y$, $Z$, and $\mu_{12}$ are shown in Figure \ref{fig:binnedPlots}.~As shown in Figure \ref{fig:massBin}, increasing the stellar mass results in a dimmer \mtrgb.~At this stage of stellar evolution, increasing the stellar mass primarily adds more mass to the envelope, leaving the core mass relatively unchanged.~This increases the opacity, resulting in a dimmer \mtrgb. Figure \ref{fig:yBin} shows that the helium abundance has only a minor effect on the brightness of the \trgb.~Figure \ref{fig:zBin}  shows that the I-band magnitude has a peak brightness at $Z\approx10^{-3}$.~This mirrors the relationship between the bolometric correction in this band and effective temperature.~The maximum amount of light is emitted into the I-band at $T_{\rm eff}\approx 4800$K, which, for our model parameters, occurs at $Z\approx 10^{-3}$.~Finally, Figure \ref{fig:muBin} shows that, as expected and explained in detail in \S\ref{sec:intro}, the brightness of the tip increases with $\mu_{12}$.

\subsection{Machine Learning}
After training the machine learning classifier with the grid to eliminate stars that helium shell flash or are  older than the age of the universe, we  found an accuracy of 99.6\% and a categorical crossentropy loss of $9.0\times10^{-3}$.~Our DNN Regressor predicts $M_I$, $(V - I)$ color, and both WL errors with a mean squared error loss of $1.4\times10^{-4}$.~The residuals for the regression algorithm are shown in Figure \ref{fig:errorHists}.The mean and standard deviation of these distributions are 1-2 orders of magnitude smaller  than, and thus subdominant to, the WL and empirical uncertainties, which are on the order of $0.1$ dex.~The  accuracy of these ML algorithms are comparable to those in \cite{Dennis:2023kfe,Dennis:2023ldw}.~

\subsection{MCMC Analysis}
\begin{figure}
    \centering
    \includegraphics[width=\columnwidth]{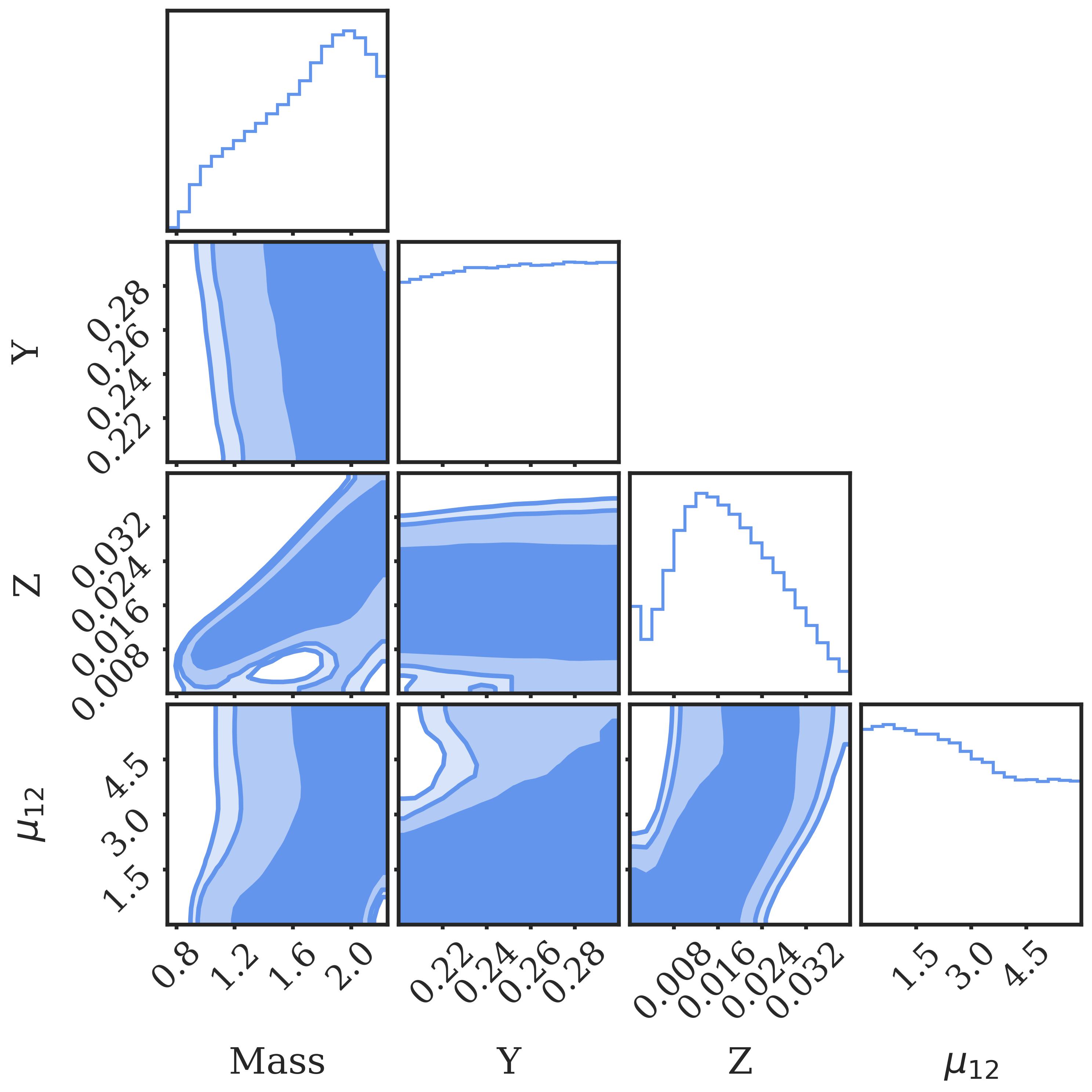}
    \caption{ Corner plot showing the results of the MCMC analysis from comparison with the \mtrgb calibration  in $\omega$-Centauri  \cite{Bellazzini2001}.~In the 2D histograms, each contour is a two dimensional confidence level with the smallest contour being 68\% and increasing in size to 90\% and 95\%.~The 1D histograms show the marginalized posterior for each input parameter.}
    \label{fig:corner_OmCent}
\end{figure}

The results from the MCMC analysis using the calibrated \mtrgb\ in $\omega$-Centauri reported by \cite{Bellazzini2001} are shown in Figure \ref{fig:corner_OmCent}.~Results of the MCMC using \mtrgb\ calibrations of in NGC 4258 \cite{JangLee} and the LMC \cite{Yuan2019} are shown in Appendix \ref{app:mcmc}.~After characterizing the uncertainties in $M$, $Y$, and $Z$, we  find that it is possible for the value of $\mu_{12}$ to fall anywhere within our parameter space because the posterior of $\mu_{12}$ is nearly flat across the entire range.~This implies that the region of parameter space $\mu_{12} \leq 6$ is  unconstrained.~This  is explained by Figure \ref{fig:zBin}, as $\mu_{12}$ increases, more models across a wider range of $Z$ and $M$ fall within $1\sigma$ of the observed \mtrgb.

\section{Discussion}
\label{sec:conclusion}
In this section we  discuss the implications of our results, suggest further applications of our method, and highlight potential caveats and limitations of our study.

\subsection{Discussion of Our Results}

The results of our MCMC are summarized in Figure \ref{fig:corner_OmCent}.~We find a broad marginalized posterior on $\mu_{12}$ spread across the entire parameter space.~We therefore conclude that the region $\mu_{12} \leq 6$ is unconstrained when the degeneracies across parameter space are accounted for.~Previous works (e.g., \cite{francesco&raffelt2020}) that do not simultaneously vary the stellar input physics with the new physics parameters find that the region $\mu_{12} < 1.2$ is excluded.~Our work highlights the importance of fully accounting for stellar uncertainties when using stars as novel probes of physics beyond the Standard Model.~

We further remark that the region $\mu_{12}>6$ is not excluded by our analysis but, rather, we are unable to explore this region since our grid does not extend to higher values.~Given our results, and the fact that we fixed several input physics parameters that are known to have large uncertainties, it is likely that large regions of parameter space with $\mu_{12}>6$ are viable once stellar uncertainties are accounted for.

The opening of the  parameter space has important implications for both stellar and terrestrial searches for a non-negligible neutrino magnetic dipole moment.~As regards stellar probes, it is possible that a large neutrino MDM could be detected using other astrophysical objects provided that degeneracies and uncertainties with stellar physics are accounted for in a similar manner to this work.~As regards to terrestrial searches, the region of parameter space we have opened is accessible to  dark matter direct detection chambers e.g., XENONnT \cite{2020PhRvD.102g2004A}, raising the possibility that these experiments could detect a non-negligible neutrino MDM.

\subsection{Comparison with Previous Works}
It is instructive to compare our results with others in the literature.~A previous work \cite{Dennis:2023kfe} (authored by two of us) used a similar method to reanalyse bounds on the axion-electron coupling, $\alpha_{26}$.~That work found similar results --- namely that a large region of the axion-electron coupling parameter space is unconstrained once stellar degeneracies and uncertainties are accounted for --- but there are differences in the final 2D probability distributions.~In this work, we found that high $\mu_{12}$ prefers high masses $M\sim2M_\odot$ and intermediate metallicities $0.008<Z<0.02$.~In contrast, \cite{Dennis:2023kfe} found that intermediate masses $M\sim 1.2M_\odot$ and $Z>0.03$ are preferred at high $\alpha_{26}$.~Such high metallicities may be unrealistic in light of other observations of the TRGB host objects we study.~

We investigated the cause of this difference and found that it can be attributed to the difference in mixing lengths adopted by the two works.~We simulated a  grid of models with $\alpha_{MLT}=1.8$ (as adopted by \cite{Dennis:2023kfe}) and  $\mu_{12}=0,\,6$.~In both cases we found a difference in the trend across parameter space between the two mixing lengths.~In particular, we found that with $\alpha_{MLT}=1.8$, there were a large number of $\mu_{12}=6$ models at intermediate mass and high metallicity that were compatible with the \mtrgb\ calibration, consistent with the results of \cite{Dennis:2023kfe}.~In contrast, Figure \ref{fig:binnedPlots} reveals that when $\alpha_{MLT}=2.0$ the largest density of compatible models occurs for high masses and intermediate metallicities.~Our investigation demonstrates that the conclusions of \cite{Dennis:2023kfe} are not reliant upon the systems studied having super-solar metallicity.~Indeed, had the mixing length been simultaneously varied  with $M$, $Y$, and $Z$ reference \cite{Dennis:2023kfe} would have found that a large region of  parameter space with high mixing length, high mass, and intermediate metallicity  --- similar to the region we found in this work --- is compatible with the data.

Our second comparison is with the results of \cite{francesco&raffelt2020}, who report the most stringent bound $\mu_\nu < 1.2 \times10^{-12} \mu_B$ at the 95\% confidence level using the calibrated \mtrgb\ in $\omega$-Centauri.~As discussed at length in \cite{Dennis:2023kfe}, the bound obtained by \cite{francesco&raffelt2020} is the result of varying stellar parameters over an extremely narrow region of parameter space that is motivated by observations of the globular cluster M5.~In the language of our methodology, this corresponds to placing tight priors on the allowed range of stellar parameters.~This bound is heavily-reliant upon the assumptions about M5 being valid e.g., that it is close to 13.8 Gyr old;~that the measurements informing the choice of priors have their errors accurately assessed --- in some cases there are competing measurements that are discrepant with one another;~and that the model for M5 is a good model for the systems considered by \cite{francesco&raffelt2020}.~In light of these  uncertainties, we advocate for our data-driven approach, although we note that if one could use similar methods to ours to assess the errors on the measurements above then it is possible that bounds could be placed provided said errors are sufficiently small.

\subsection{Application to Other Stellar Probes of the Neutrino Magnetic Dipole Moment}
An important application of our method is to reevaluate other stellar bounds on the neutrino MDM.~There are no astrophysical probes of fundamental physics that are free from degeneracies and uncertainties, and our work has highlighted the importance of accounting for these when placing  bounds.~In the case of the neutrino MDM, our methodology could be used to revisit bounds coming from the Sun \cite{Borexino:2017fbd} and the white dwarf luminosity function \cite{Bertolami2014}.~With the TRGB bounds now allowing values of $\mu_{12}\leq6$, the latter test, which yields $\mu_{12} < 5$, is presently the most constraining.

\subsection{Limitations of Our Study}
\label{sec:limits}

Our work is subject to some limitations that we now discuss.~First, as a result of limited computation time, we do not vary all possible input parameters such as mixing length, radiative opacity, conductive opacity, nuclear reaction rates, neutrino loss rate, mass loss, and others listed in \cite{serenelli2017,Ippocratis2022}.~As a result, our uncertainties are most likely underestimated.~In future work, this could be improved upon by varying more parameters in a larger grid, increasing the number of stellar simulations.~This would be computationally expensive.~For example, if we were to vary 10 stellar parameters the grid would  take $\sim$2 billion CPU years to complete.~Our grid with four parameters varying completed in $\sim$23 CPU years.~More efficient sampling methods such as Latin hypercube sampling or active learning \cite{2022arXiv220316683A} could speed this up dramatically.~Another novel challenge in this endeavor is the longer convergence times for random walk MCMC algorithms in higher-dimensional parameter spaces.~This could be overcome by using Hamiltonian MCMC (HMCMC).~A major advantage of using DNNs is that they are differentiable, which is necessary for HMCMC.~We remark that increasing further sources of uncertainty can only strengthen the conclusions of this work, but doing so would allow one to determine the boundary of the region of parameter space excluded by TRGB measurements.

A second limitation is that we define the \trgb\ using a single stellar evolution model.~This is tantamount to assuming that the TRGB I-band magnitude is due to the brightest star.~The goal of this work is to compare with other works that made this identical assumption (e.g.~\citep{Raffelt1995, Capozzi2020}).~Observationally, \mtrgb\ is calibrated using edge-finding techniques applied to the  entire color-magnitude diagram \cite{Lee1993}, which includes a multitude of stars with varying stellar parameters and a complex star formation history.~Going beyond the single-star approximation would enable a more consistent comparison with observation.~One could accomplish this by  first using our ML emulator to create synthetic color-magnitude diagrams  by simulating a population of stars with parameters drawn from some distribution, and  then applying the same edge-finding techniques  to these to obtain a more realistic prediction for \mtrgb.~Using edge-finding techniques would increase the uncertainties and, as a consequence, increase the unconstrained region because there would be a larger spread in the simulated values of \mtrgb.~This makes an interesting topic for future work.

A final limitation is that our method includes statistical uncertainties but not systematic uncertainties.~For example, we used a single stellar structure code with certain discrete choices e.g., elemental abundances, numerical solver, atmosphere model, etc.~Using a different code may give different results.~Similarly, important physics not captured by MESA e.g., three-dimensional processes could be another source of systematics.~Finally, we used the bolometric corrections of Worthey \& Lee \cite{worthey&lee2011} but other choices e.g., MARCS \cite{2008A&A...486..951G} or PHOENIX \cite{2008ApJS..178...89D} may give different predictions for \mtrgb\ and its empirical uncertainty.~It would be interesting to repeat our work using  different stellar structure codes and different bolometric corrections and compare the results.

\section{Conclusions}
\label{sec:real_conclusions}

In this work we utilized a novel methodology to constrain the neutrino magnetic dipole moment (MDM) using empirical calibrations of the tip of the red giant branch (TRGB) I-band magnitude (\mtrgb) that fully accounts for uncertainties and degeneracies due to stellar input physics.~We first simulated 146,250 MESA models, varying the stellar mass ($M$), helium abundance ($Y$), metallicity ($Z$), and the reduced neutrino MDM ($\mu_{12} = \mu_\nu \times 10^{-12}~\mu_{B}$).~We then used the bolometric correction code from Worthey \& Lee \cite{worthey&lee2011} to calculate the I-Band magnitude of the \trgb\ for each model.~Next, we trained a machine learning algorithm to predict \mtrgb\ and its associated uncertainty given values of the four input parameters.~Finally, we used  our ML emulator to make theoretical predictions for \mtrgb\ in an MCMC analysis  that compared them with empirical calibrations to constrain the neutrino MDM.
We found that once stellar uncertainties are fully accounted for, observations of the \trgb\ do not place any constraints on the neutrino MDM when $\mu_{12}\leq 6$.~Furthermore, given the broad and flat posterior we found, it is likely that larger values are equally unconstrained, although our grid does not extend to larger values to confirm this.~Our study has opened up a large region
of parameter space that was previously believed to be excluded.~This region is accessible to planned terrestrial dark matter direct detection chambers.~It is highly likely that the bounds on the neutrino MDM from other stellar probes will be similarly weakened once
degeneracies are fully accounted for.~The methodology we
have utilized can be applied to these tests in a straightforward manner, and doing so is of paramount importance
for determining the viable parameter space of theories
beyond the Standard Model.

\section{Software}
MESA version 12778, MESASDK version 20.3.2, python version 3.8, \texttt{NumPy} version 1.22.3 \citep{NumPy}, \texttt{Pandas} version 1.4.3 \citep{pandasDataStructure, pandasSoftware}, \texttt{Matplotlib} version 3.5.1 \cite{matplotlib}, \texttt{Seaborn} version 0.11.2 \cite{Seaborn},  \texttt{Tensorflow} version 2.4.1 \citep{tf, keras}, \texttt{corner} version 2.2.1 \cite{corner}, \texttt{emcee} version 3.1.2 \citep{emcee}.

\section{Acknowledgements}
We thank Adrian Ayala, Aaron Dotter, Robert Farmer, Frank Timmes, and the wider MESA community for answering our MESA-related questions.~We are grateful for conversations with Eric J.~Baxter, Djuna Croon, Samuel D.~McDermott, Harry Desmond, Marco Gatti, Dan Hey, Jason Kumar, Danny Marfatia, Marco Raveri, David Rubin, Xerxes Tata, and Guy Worthey.~We are especially thankful to David Schanzenbach for his assistance with using the University of Hawai\okina i MANA supercomputer.

NF acknowledges support from Research Experience for Undergraduate program at the Institute for Astronomy, University of Hawai'i-Manoa funded through NSF grant \#2050710.~NF would like to thank the Institute for Astronomy for their hospitality during the course of this project.~To run our stellar evolution simulations and much of our analysis we utilized the University of Hawai\okina i’s high-performance supercomputer MANA.~The technical support and advanced computing resources from University of Hawaii Information Technology Services – Cyberinfrastructure, funded in part by the National Science Foundation MRI award \#1920304, are gratefully acknowledged.

\bibliography{main}

\appendix

\section{Further MCMC Results}\label{app:mcmc}
\begin{figure}
    \centering
    \includegraphics[width=\columnwidth]{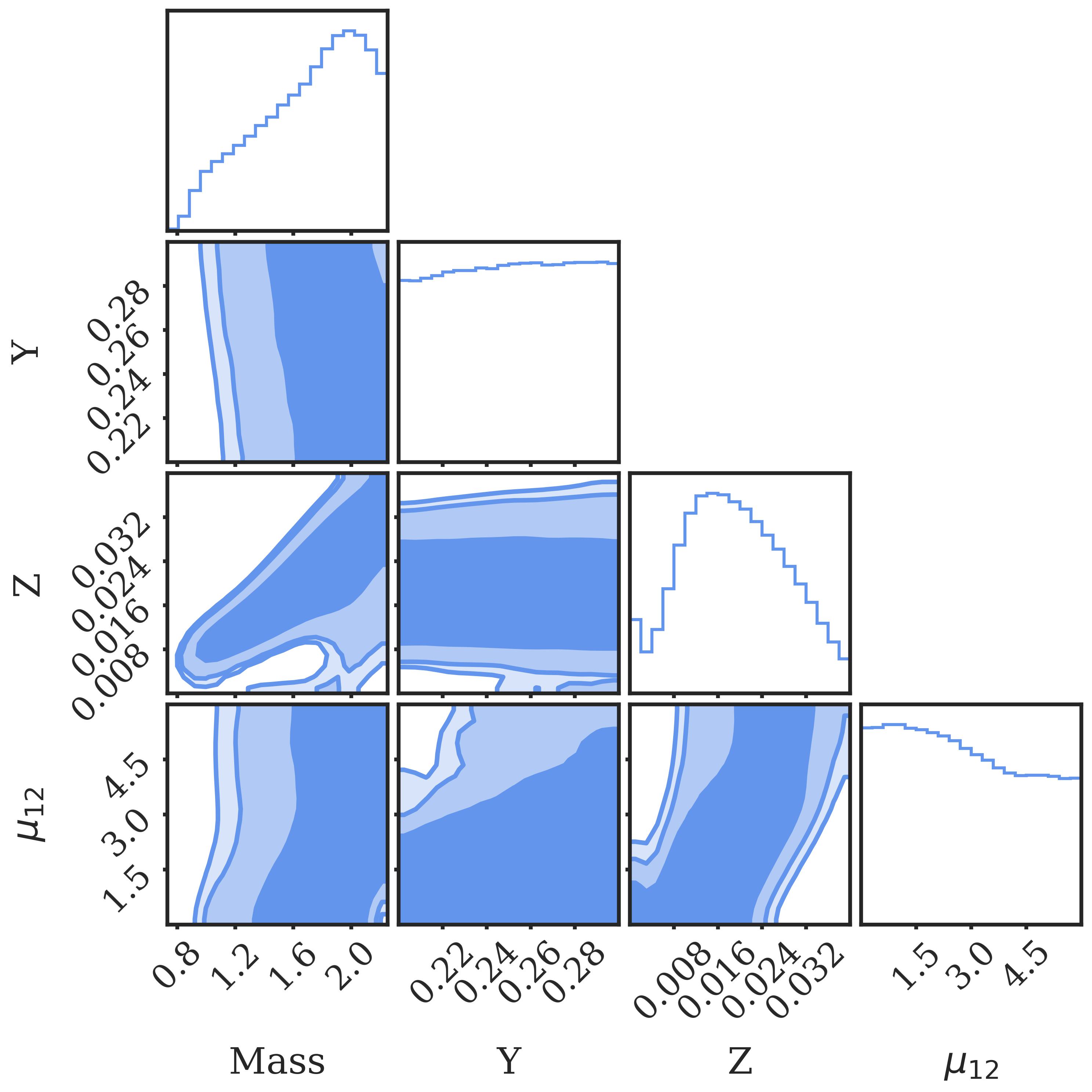}
    \caption{Same as Figure \ref{fig:corner_OmCent} except the MCMC analysis was done with comparison to the \mtrgb\ calibration from \cite{Yuan2019}.}
    \label{fig:corner}
\end{figure}

\begin{figure}
    \centering
    \includegraphics[width=\columnwidth]{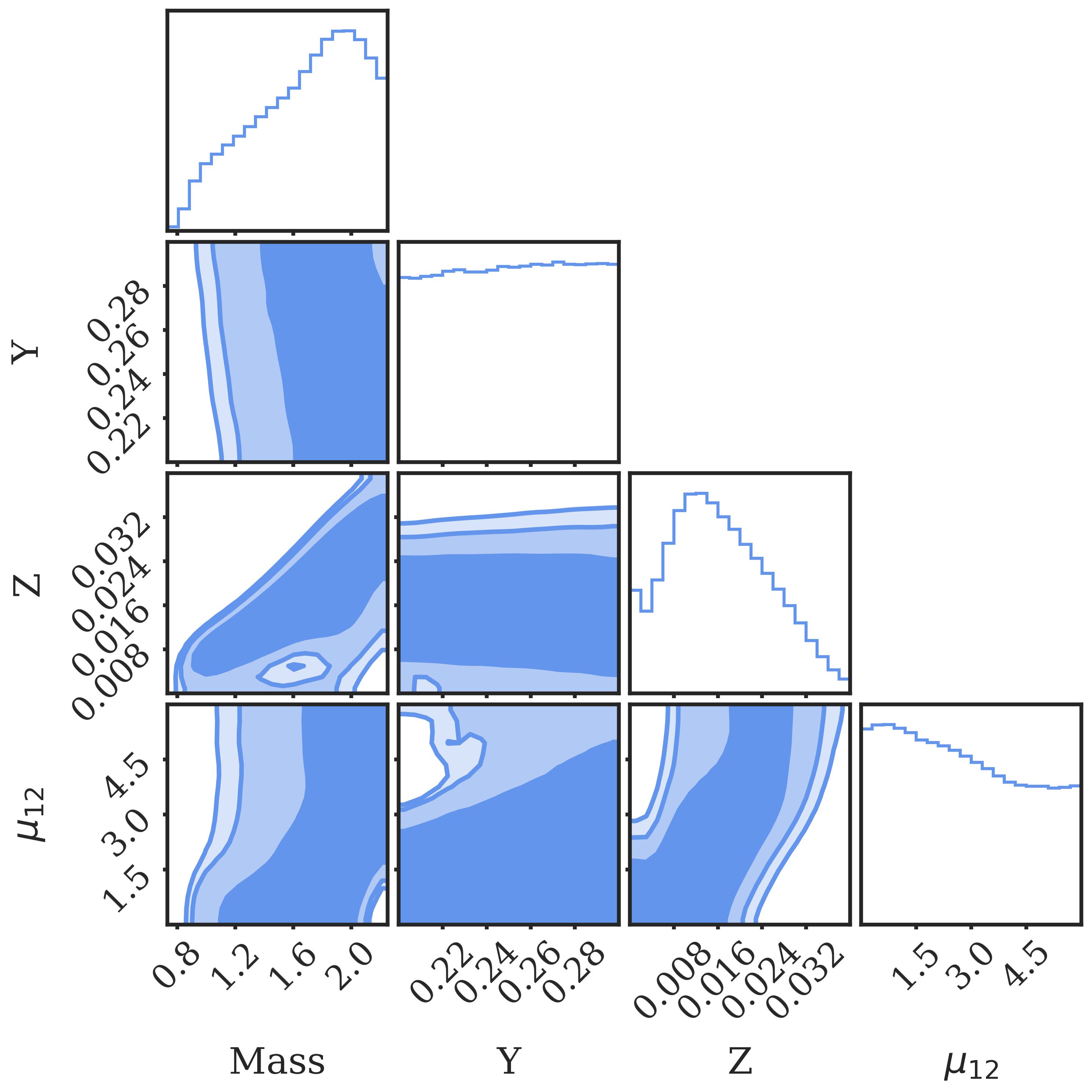}
    \caption{Same as Figure \ref{fig:corner_OmCent} except the MCMC analysis was done with comparison to the \mtrgb\ calibration from \cite{JangLee}.}
    \label{fig:corner_NGC}
\end{figure}

Corner plots showing the results from comparison with the \mtrgb calibrations from \cite{Yuan2019} and \cite{JangLee} are in Figures \ref{fig:corner} and \ref{fig:corner_NGC}, respectively.

\clearpage

\end{document}